\documentclass[twocolumn,showpacs,floatfix,prl,reprint]{revtex4-1}

\usepackage{graphicx}
\usepackage{amssymb}
\usepackage{amsmath}
\usepackage{color}
\usepackage{psfrag}
\usepackage{epsfig}
\usepackage{bbm}
\usepackage{bm}
\usepackage{dsfont}

\usepackage{hyperref}
\hypersetup{breaklinks}

\newcommand{\ket}[1]{| #1 \rangle}

\newcommand{\rb}[1]{\left( #1 \right)}

\newcommand{\ew}[1]{\langle #1 \rangle}
\newcommand{\beq}{\begin{eqnarray}}
\newcommand{\eeq}{\end{eqnarray}}

\newcommand{\op}[2]{| #1 \rangle \langle #2 |}

\newcommand{\eq}[1]{Eq.~(\ref{#1})}
\newcommand{\fig}[1]{Fig.~\ref{#1}}

\newcommand{\citer}[1]{{Ref.~\cite{#1}}}

\begin{document}
\title{Dark states in multi-mode multi-atom Jaynes-Cummings systems}

\author{C.~Emary}
\affiliation{
  Department of Physics and Mathematics,
  University of Hull,
  Kingston-upon-Hull,
  HU6 7RX,
  United Kingdom}

\date{\today}
\begin{abstract}
  We consider a system consisting of $N$ two-level atoms inside an $M$-mode  degenerate, driven cavity.   We discuss the formation of dark states in this system and derive the conditions required for the observation of the dark-state anti-resonance in the cavity emission.
\end{abstract}
\maketitle
%%%%%%%%%%%%%%%%%%%%%%%%%%%%%%%%%%%%%%%%%%%%%%%%%%%%%%%%%%%%%%%%%%%%%%%%%

%%%%%%%%%%%%%%%%%%%%%%%%%%%%%%%%%%%%%%%%%%%%%%%%%%%%%%%%%%%%%%%%%%%%%%%%%
%%%%%%%%%%%%%%%%%%%%%%%%%%        INTRO        %%%%%%%%%%%%%%%%%%%%%%%%%%
%%%%%%%%%%%%%%%%%%%%%%%%%%%%%%%%%%%%%%%%%%%%%%%%%%%%%%%%%%%%%%%%%%%%%%%%%

Dark states are superposition states which, in quantum optics at least, decouple from the relevant light field(s) due to the interference between transitions originating from the different elements of the superposition.  They lie of the heart of many quantum-optical phenomena such as coherent population trapping \cite{Alzetta1976,Arimondo1976,Whitley1976,Arimondo1996}, electromagnetically induced transparency \cite{Fleischhauer2005}, amplification without inversion \cite{Kocharovskaya1992}, lasing without inversion \cite{Mompart2000} and STIRAP \cite{Bergmann1998}.  
They are also of interest further-a-field in areas such as quantum transport \cite{Brandes2000,Michaelis2006,Groth2006,Emary2007,Poeltl2009,Busl2012,Poeltl2012} or quantum information processing \cite{Faoro2003,Troiani2003,Kis2004,Paspalakis2004,Greentree2004,Chen2004,Emary2007a}.

The prototypical system in which dark states occur is a three-level system in the $\Lambda$-configuration where each of the ground-to-excited-state transitions is driven coherently by monochromatic classical fields.  
In contrast, the Jaynes-Cummings model (JCM) \cite{Jaynes1963} consists of a {\em two-level} system interacting with a single {\em quantised} field mode.  Nevertheless,  dark states can also occur in the JCM provided that an additional external coherent driving is present.
Cirac \cite{Cirac1993}, building on the work of \cite{Alsing1992} and in the
context of ion traps, showed that classically driving the atom could drive the JCM into a state with the atom in its ground state and field in a squeezed state. Atomic fluorescence is thus suppressed and, in this sense, the system becomes dark.  This effect was later dubbed ``cavity-induced transparency'' in \citer{Rice1996} and understood in terms of interference between the cavity field and atomic polarisation.  The effect of these dark states arising from atomic driving in narrowing line shapes was discussed in \citer{Zippilli2004} for a system of $N \ge 1$ atoms.
Instead of driving the atom(s), the cavity mode can be directly driven by the external laser.  This can drive the system into a state which is approximately dark in the single-atom case \cite{Peano2010} and which
approaches complete darkness in the limit that the number of atoms $N$ becomes large \cite{Milburn2000}.

In this work, we consider a system consisting of $N$ two-level atoms inside an degenerate $M$-mode coherently-driven cavity, as was recently realised in the experiment of \citer{Wickenbrock2013}.  We discuss the formation of dark states in this system.  These states are superpositions of the atomic ground-state with a collective atomic state where a single excitation delocalised across all the atoms.  We describe the conditions required for the observation of dark-state anti-resonance in the  emission from the multi-mode cavity.

%%%%%%%%%%%%%%%%%%%%%%%%%%%%%%%%%%%%%%%%%%%%%%%%%%%%%%%%%%%%%%%%%%%%%%%%
%%%%%%%%%%%%%%%%%%%%%%%%%       Model       %%%%%%%%%%%%%%%%%%%%%%%%%%%%
%%%%%%%%%%%%%%%%%%%%%%%%%%%%%%%%%%%%%%%%%%%%%%%%%%%%%%%%%%%%%%%%%%%%%%%%
\section{Atom-cavity system}

We consider an $N$-atom JCM (or Tavis-Cummings model \cite{Tavis1968}) extended to include $M$ cavity modes and an additional coherent monochromatic driving term \cite{Wickenbrock2013}.  The Hamiltonian reads
\beq
  H = H_S + H_D 
  \label{Hfull}
  ,
\eeq
where $H_S$ describes the system and $H_D$, the driving.  In a frame rotating with the driving frequency, $H_S$ reads ($\hbar=1$ throughout)
\beq
  H_S &=& 
  -\Delta_C \sum_{k=1}^M a_k^\dag a_k
  -\frac{1}{2} \Delta_A \sum_{l=1}^N \sigma_l^z
  \nonumber\\
  &&
  +
  \frac{1}{2}
  \sum_{k=1}^M 
  \sum_{l=1}^N 
  \left\{
    g_{kl} a_k^\dag \sigma^-_l
    +
    g^*_{kl} a_k  \sigma^+_l
  \right\}
  ,
\eeq
where $\Delta_C$ and $\Delta_A$ are the detunings of the cavity and atomic transitions respectively (assumed homogeneous), and $g_{kl}$ is the coupling constant for the interaction between the $k$th mode and the $l$th atom. In the rotating frame, the driving of the cavity is described by the Hamiltonian
\beq
   H_D &=& 
  \sum_{k=1}^M \eta_k 
  (a_k^\dag + a_k )
  ,
\eeq
with $\eta_k$ the coupling strength of the laser into mode $k$.

To include the effects of cavity losses and spontaneous emission we study the
Liouville-von-Neumann equation for the atom-cavity density matrix $\rho$, which reads
\beq
  \frac{d}{dt}\rho = -i \left[H, \rho\right] 
  +\mathcal{L}_\mathrm{loss}[\rho]
  +  \mathcal{L}_\mathrm{spon}[\rho]
  ,
\eeq
with
\beq
  \mathcal{L}_\mathrm{loss}[\rho] = 
  \frac{\kappa}{2}\sum_{m=1}^M 
  \left\{
    2 a_m \rho a_m^\dag 
    - a_m^\dag a_m \rho 
    - \rho a_m^\dag a_m  
  \right\}
  \label{Lloss}
  ,
\eeq
describing cavity loss at rate $\kappa$, and
\beq
  \mathcal{L}_\mathrm{spon}[\rho] = 
  \frac{\gamma}{2}\sum_{n=1}^N 
  \left\{
    2 \sigma_n^- \rho \sigma_n^+ 
    - \sigma_n^+ \sigma_n^- \rho 
    - \rho \sigma_n^+ \sigma_n^-
  \right\}
  \label{Lspon}
  ,
\eeq
describing spontaneous emission from the atoms at rate $\gamma$. We have assumed homogeneous rates here for simplicity.

%%%%%%%%%%%%%%%%%%%%%%%%%%%%%%%%%%%%%%%%%%%%%%%%%%%%%%%%%%%%%%%%%%%%%%%%
%%%%%%%%%%%%%%%%%%%%%%          WEAK         %%%%%%%%%%%%%%%%%%%%%%%%%
%%%%%%%%%%%%%%%%%%%%%%%%%%%%%%%%%%%%%%%%%%%%%%%%%%%%%%%%%%%%%%%%%%%%%%%%
\section{Weak excitation model }
We consider that the effect of the driving term $H_D$ is weak enough that it introduces at most one excitation (atomic flip or cavity photon) into the system at any given time.  This assumption allows us to restrict ourselves to the following set of relevant states:
\beq
  \ket{\underline{0}} &:& \quad \text{cavity vacuum and all atoms in ground state} \nonumber\\
  \ket{C_k} &:& \quad \text{single-photon excitation in cavity mode}~k\nonumber\\
  \ket{A_l} &:& \quad \text{excitation of atom}~l \nonumber
\eeq
In this restricted basis, our Hamiltonian (minus an offset) read
\beq
  \label{weakmodel}
  H &=& 
  - \Delta_C \sum_{k=1}^M \op{C_k}{C_k}
  -\Delta_A \sum_{l=1}^N  \op{A_l}{A_l}
  \nonumber\\
  &&+
  \frac{1}{2}\sum_{k=1}^M \sum_{l=1}^N
  \rb{
    g_{kl} \op{C_k}{A_l} +g^*_{kl}  \op{A_l}{C_k}
  }
  \nonumber
  \\
  &&
  +
  \sum_{k=1}^M \eta_k \rb{ \op{C_k}{\underline{0}} + \op{\underline{0}}{C_k}}  
  .
\eeq
The $M\times N$ coupling-constant matrix $G$ with elements $(G)_{kl}=g_{kl}$ has the singular value decomposition $G = U \Lambda W^\dag$, with $U$ an $M\times M$ unitary matrix, $W$ an $N\times N$ unitary matrix, and $\Lambda$ an $M\times N$ rectangular diagonal matrix.  The  non-zero diagonal elements of $\Lambda$ are the singular values of $G$, $\lambda_j$, which are real, positive and equal to the non-zero eigenvalues of the matrix  $\Gamma \equiv \sqrt{G G^\dag}$.  The number of singular values is equal to the rank,  $R_\Gamma$, of this matrix.  We shall assume that $M \le N$, such that the coupling matrix $G$ can have at most $M$ singular values: $R_\Gamma \le M$.

We then define the {\em collective} cavity and atomic states:
\beq
   \ket{\widetilde{C}_j} &=& \sum_{k=1}^M  U_{kj} \ket{C_k} \quad \quad \mathrm{(cavity)}
   \label{colstateC}\\
    \ket{\widetilde{A}_j} &=& \sum_{l=1}^N W_{lj}\ket{A_l}\quad \quad \mathrm{(atomic)}
    \label{colstateA}
\eeq 
to rewrite the Hamiltonian as
\beq
  H =  h_0 +  \sum_{j=1}^{R_\Gamma} h_j + H_D
  \label{Hsplit}
  ,
\eeq
with an uncoupled component arising from a rank-deficient $\Gamma$-matrix,
\beq
  h_0 = \sum_{j= R_\Gamma+1}^M
  -\Delta_C \op{\widetilde{C}_j}{\widetilde{C}_j}
  - 
  \Delta_A \op{\widetilde{A}_j}{\widetilde{A}_j}
  ,
\eeq
and the $R_\Gamma$ components describing the  coupling between the $j$th collective atom and cavity modes,
\beq
  h_j &= &
    -\Delta_C \op{\widetilde{C}_j}{\widetilde{C}_j}
    - \Delta_A  \op{\widetilde{A}_j}{\widetilde{A}_j}
    \nonumber\\
    &&
    + \frac{1}{2}  \lambda_j 
    \rb{
      \op{\widetilde{C}_j}{\widetilde{A}_j} 
      +
      \op{\widetilde{A}_j}{\widetilde{C}_j}
    }
  .
\eeq 
The driving term in this collective basis reads
\beq
  H_D &=& 
  \sum_{j=1}^M  \rb{\tilde{\eta}_j \op{\widetilde{C}_j}{\underline{0}} +\tilde{\eta}^*_j \op{\underline{0}}{\widetilde{C}_j}}  
  ,
\eeq
with the new amplitudes 
$
  \tilde{\eta}_j =\sum_{k=1}^M U_{jk}^*\eta_k
$.
Thus, since the singular value decomposition mixes all cavity modes for a generic coupling matrix, pumping just one of the original modes pumps all the collective modes.

Finally, in this basis the dissipative parts of the Liouville-von-Neumann equation are of the standard form (as in \eq{Lloss} and \eq{Lspon}) with jump operators $\op{\underline{0}}{\widetilde{C}_k}$ for the cavity excitations (at rate $\kappa$) and with $\op{\underline{0}}{\widetilde{A}_l}$ for the atoms (at rate $\gamma$).

%%%%%%%%%%%%%%%%%%%%%%%%%%%%%%%%%%%%%%%%%%%%%%%%%%%%%%%%%%%%%%%%%%%%%%%%%
%%%%%%%%%%%%%%%%%%%%%      Dark states    %%%%%%%%%%%%%%%%%%%%%%%%%%%%%%%
%%%%%%%%%%%%%%%%%%%%%%%%%%%%%%%%%%%%%%%%%%%%%%%%%%%%%%%%%%%%%%%%%%%%%%%%%
\section{Dark states}

\subsection{Single cavity mode}
We first consider the case of a single cavity mode, as this illustrates many of the important features of interest here.  In this case, there is a single $\lambda$ value, which, assuming that all the atoms couple approximately equally to the mode, is given by $\lambda \sim g \sqrt{N}$, where $g$ is the coupling strength of a single atom.  We assume the driving strength $\eta$ (we drop the subscripts for this single mode case) to be real.  For simplicity, we shall also set $\Delta_A=\Delta_C=\Delta$ throughout (atom-cavity resonance).

%%%%%%%%%%%%%%%%%%%%%%%%%%%%%%%%%%%%%%%%%%%%%%%%%%%%%%%%%%%%%%%%%
\begin{figure}[tb]
  \begin{center}
    \includegraphics[width=\columnwidth,clip]{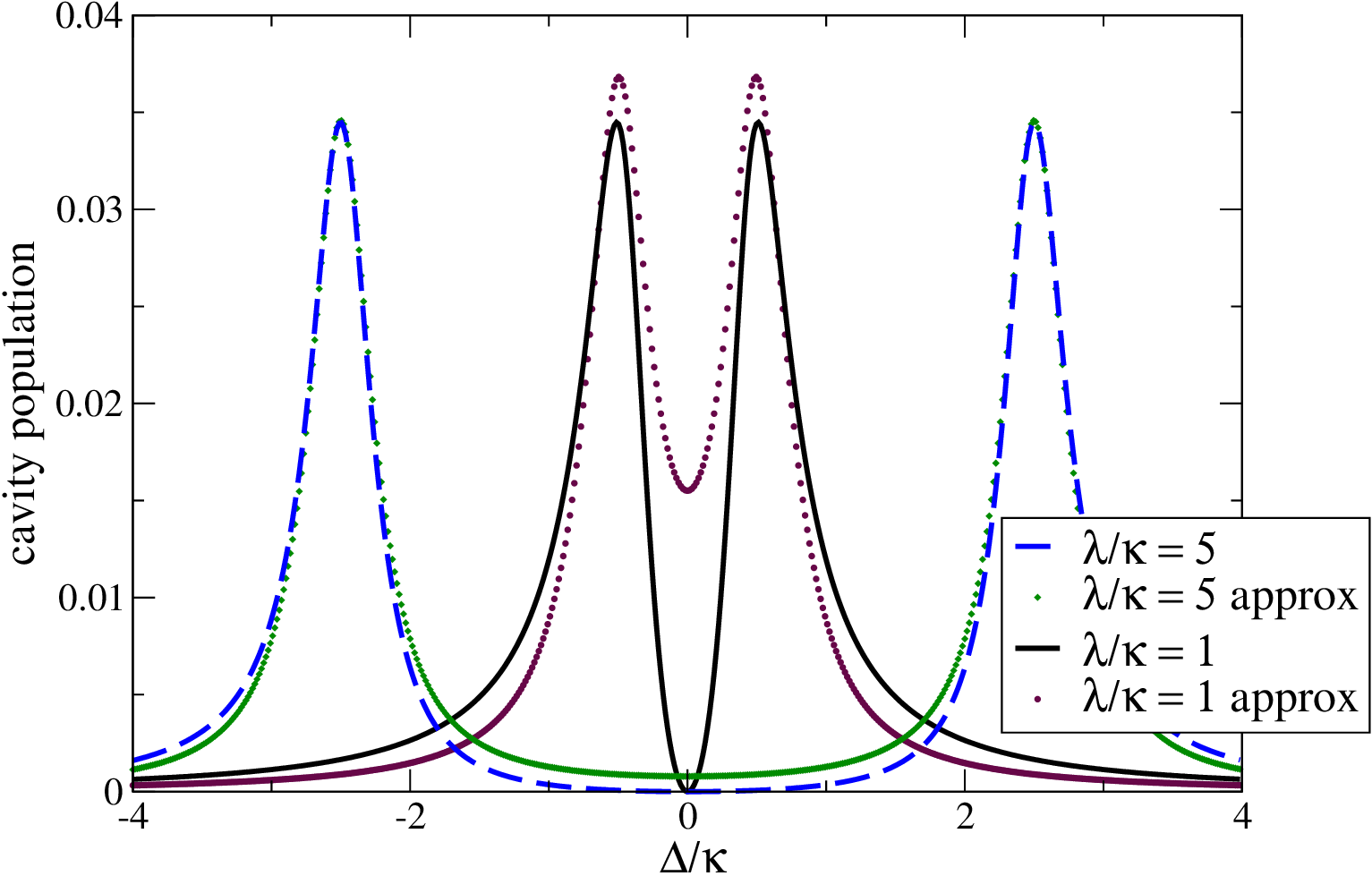}
    \caption{ 
    Stationary cavity population as a function of detuning $\Delta$ for two values of the level splitting in the single-mode case ($M=1$).  The spontaneous emission rate $\gamma$ was zero here and energies are measured in units of the cavity loss rate $\kappa$.  The black-continuous and blue-dashed lines plot the population calculated within the weak-excitation limit. The small symbols show the results from the approximation of \eq{twopeaks}.
    For $\lambda/\kappa = 5$, the population is well described by two independent Lorentzians. However, for $\lambda/\kappa = 1$ this approximation breaks down around the origin ($\Delta=0$) and a sharp anti-resonance develops. This anti-resonance is the consequence of dark-state formation.  The driving strength was $\eta/\kappa = 0.1$.
    \label{FIG:DSgam0}
    }
  \end{center}
\end{figure}
%%%%%%%%%%%%%%%%%%%%%%%%%%%%%%%%%%%%%%%%%%%%%%%%%%%%%%%%%%%%%%%%%

We will concentrate on the stationary cavity population $\ew{a^\dag a} = \mathrm{Tr} \left\{\op{C}{C} \rho_\mathrm{stat}\right\}$, with $\rho_\mathrm{stat}$ the stationary density matrix, as the cavity emission is proportional to this quantity.  
In \fig{FIG:DSgam0} we show the cavity population for two example sets of parameters.  With a large number of atoms, we can arrange that $\lambda \gg \gamma,\kappa,|\tilde\eta|$.  In this case, the emission is essentially a sum of two Lorentzians centred at $\Delta_A = \pm \lambda/2$:
\beq
  \ew{a^\dag a} \approx 
  \sum_\pm \frac{4\eta^2}{16(\Delta \pm \lambda/2)^2 + 16 \eta^2 + (\gamma + \kappa)^2}
  \label{twopeaks}
  .
\eeq
This approximation is also shown in \fig{FIG:DSgam0} and when the two peaks are well separated ($\lambda/\kappa = 5$ in the figure), it works very well.  
As the splitting decreases we would expect these two peaks to merge at around $\lambda \sim \kappa$ such that a finite population at $\Delta =0$ develops.
This does not happen, however, but rather a sharp anti-resonance emerges and the population at zero-detuning remains completely suppressed.  This anti-resonance is characteristic of the formation of a dark state.

With zero detuning, $\Delta = 0$, and spontaneous emission rate, $\gamma=0$, it is simple to verify that that state
\beq
  \ket{\Psi_D} =\mathcal{N}^{-1}
    \left\{
    \ket{\underline{0}}
    - \frac{2\eta}{\lambda}\ket{\widetilde{A}_1}
  \right\}
  ,
\eeq
with norm $\mathcal{N}^2 = 1 +4 \frac{\widetilde{\eta}^2}{\lambda^2} $,
is an exact eigenstate of the  Hamiltonian, \eq{Hsplit}, with eigenvalue zero.  Since this state has a completely empty cavity, emission will be exactly zero and the state $\ket{\Psi_D}$ is therefore ``dark''.  This darkness arises from destructive interference between the paths $\ket{\underline{0}} \to \ket{C_k}$ and $\ket{A_k}  \to \ket{C_k}$.  As can seen by solution of the master equation, cavity losses drive the system into this state, such that it represents the stationary state of the system.

At finite detuning (but still with $\gamma=0$), the stationary population of the cavity is
\beq
  \ew{a^\dag a}_{\gamma=0} = 
  \frac{16 \Delta^2 \eta^2}{16 \Delta^4 + 4 \Delta^2(8\eta^2 + \kappa^2 - 2 \lambda^2)+(4\eta^2 + \lambda^2)^2}
  ,
  \nonumber\\
  \label{cavpopgam0}
\eeq
which vanishes for $\Delta \to 0$.  From this result, we determine that for the
the anti-resonance to be distinguishable from the regular mode-splitting requires that
\beq
  \lambda^2 \lesssim \kappa^2 + 16 \eta^2
  .
  \label{DScond1}
\eeq
When this is fulfilled, the width of the anti-resonance can be approximated as
\beq
  w_\mathrm{dark}\sim \frac{4 \eta^2 + \lambda^2}{\sqrt{2\rb{16 \eta^2 + \kappa^2}}}
  .
\eeq

%%%%%%%%%%%%%%%%%%%%%%%%%%%%%%%%%%%%%%%%%%%%%%%%%%%%%%%%%%%%%%%%%
\begin{figure}[tb]
  \begin{center}
    \includegraphics[width=\columnwidth,clip]{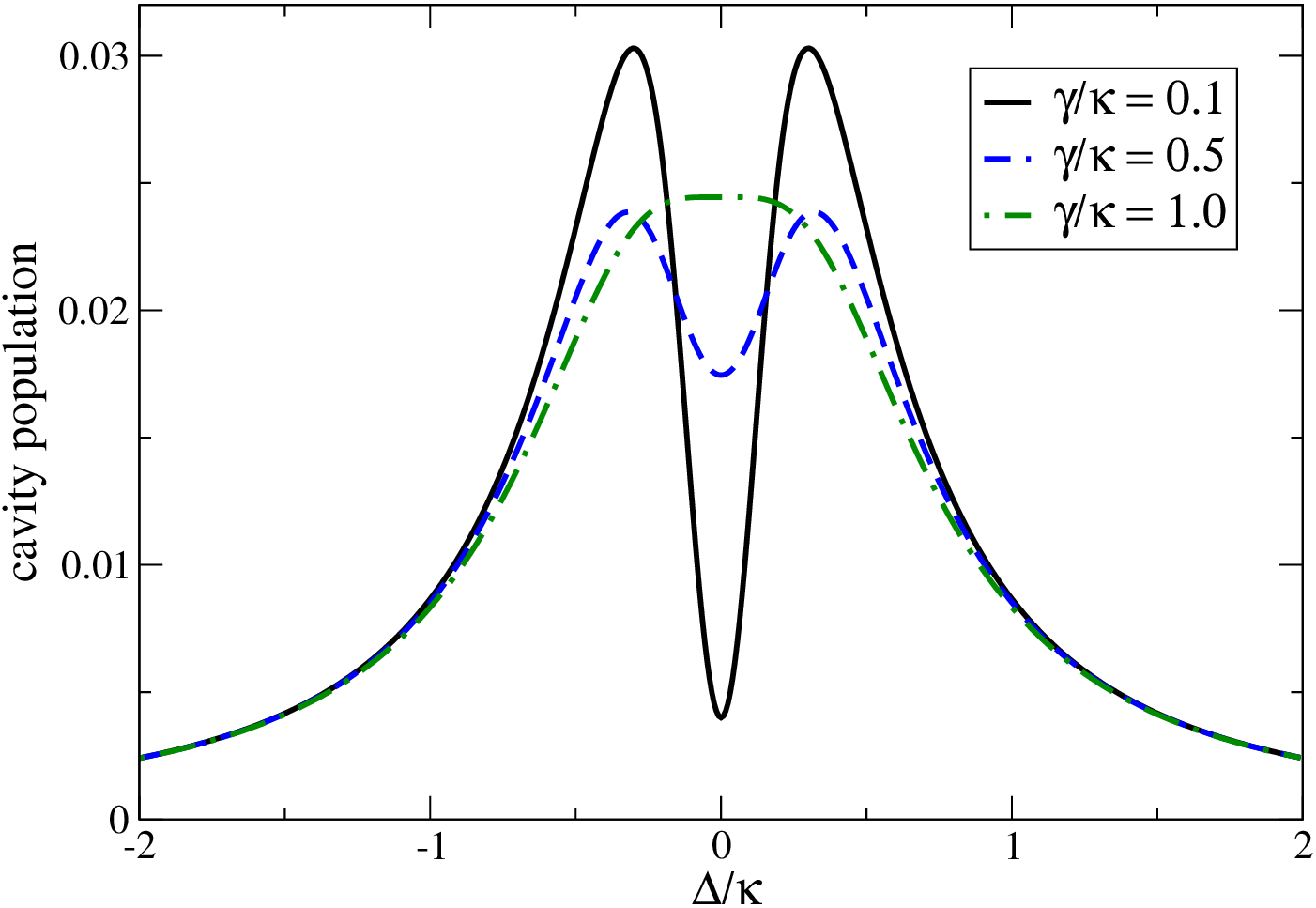}
    \caption{ 
    Stationary cavity population as a function of detuning in the $M=1$ case. Results for three values of the spontaneous emission rate $\gamma$ are shown. As $\gamma$ increases, the dark anti-resonance disappears. Here $\lambda/\kappa = 1/2$ and $\eta/\kappa = 0.1$.
    \label{FIG:DSgamfinite}
    }
  \end{center}
\end{figure}
%%%%%%%%%%%%%%%%%%%%%%%%%%%%%%%%%%%%%%%%%%%%%%%%%%%%%%%%%%%%%%%%%

The presence of spontaneous emission destroys the dark state, as can be observed from  \fig{FIG:DSgamfinite}.  The extent of this effect can be assessed by considering the population at $\Delta=0$ as a function of $\gamma$:
\beq
  \ew{a^\dag a}_{\Delta=0} &=& 
  \left\{\frac{}{}
    4 \eta^2 \gamma \left[4 \eta^2 + \gamma(\gamma+\kappa)\right]
  \right\}
  \times
  \\
  &&
%   \times
  \left\{\frac{}{}
    \gamma(8 \eta^2 + \kappa^2)\left[4 \eta^2 + \gamma(\gamma+\kappa)\right]
  \right.
  \nonumber\\
  &&
  \left.
    +
    2\kappa\lambda^2\left[2 \eta^2 + \gamma(\gamma+\kappa)\right]
    +
    (\gamma+\kappa)\lambda^4
  \frac{}{}\right\}^{-1}
  \nonumber
  \label{popDelta0}
  .
\eeq
From this, it is apparent that a significant suppression of the population at $\Delta =0$ only occurs if the splitting obeys
\beq
  \lambda^2 \gg \frac{2\gamma}{\kappa}\rb{8\eta^2 + \kappa^2}
  \label{DScond2}
  .
\eeq
This suppression may be due to either the dark state formation or simply because of the peaks being far apart.  The latter possibility is excluded, however, by the condition \eq{DScond1}.  Taking these two conditions together, and making use of 
the fact that the weak-assumption limit implies $\eta \ll \kappa$, the condition for the observability of the dark anti-resonance reads:
\beq
  \sqrt{2 \gamma \kappa} \ll \lambda \lesssim \kappa
  .
\eeq
From this it is clear that, independent of $\lambda$, the dark anti-resonance can only occur when $2\gamma/\kappa \ll 1$, i.e. when the spontaneous emission rate is significantly smaller than the cavity losses.  In the experiment of \citer{Wickenbrock2013}, this figure was $2\gamma/\kappa \approx 6.5$, so the dark-state anti-resonance would not be observable without opening the cavity somewhat. Assuming  twenty-fold increase in $\kappa$, then with $g/\kappa =0.12/(20\times 0.8)=0.0075$, the number of atoms required to achieve $\lambda \sim \kappa$ is of the order $10^4$, which is within the experimental range.

%%%%%%%%%%%%%%%%%%%%%%%%%%%%%%%%%%%%%%%%%%%%%%%%%%%%%%%%%%%%%%%%%
\begin{figure}[tb]
  \begin{center}
    \includegraphics[width=\columnwidth,clip]{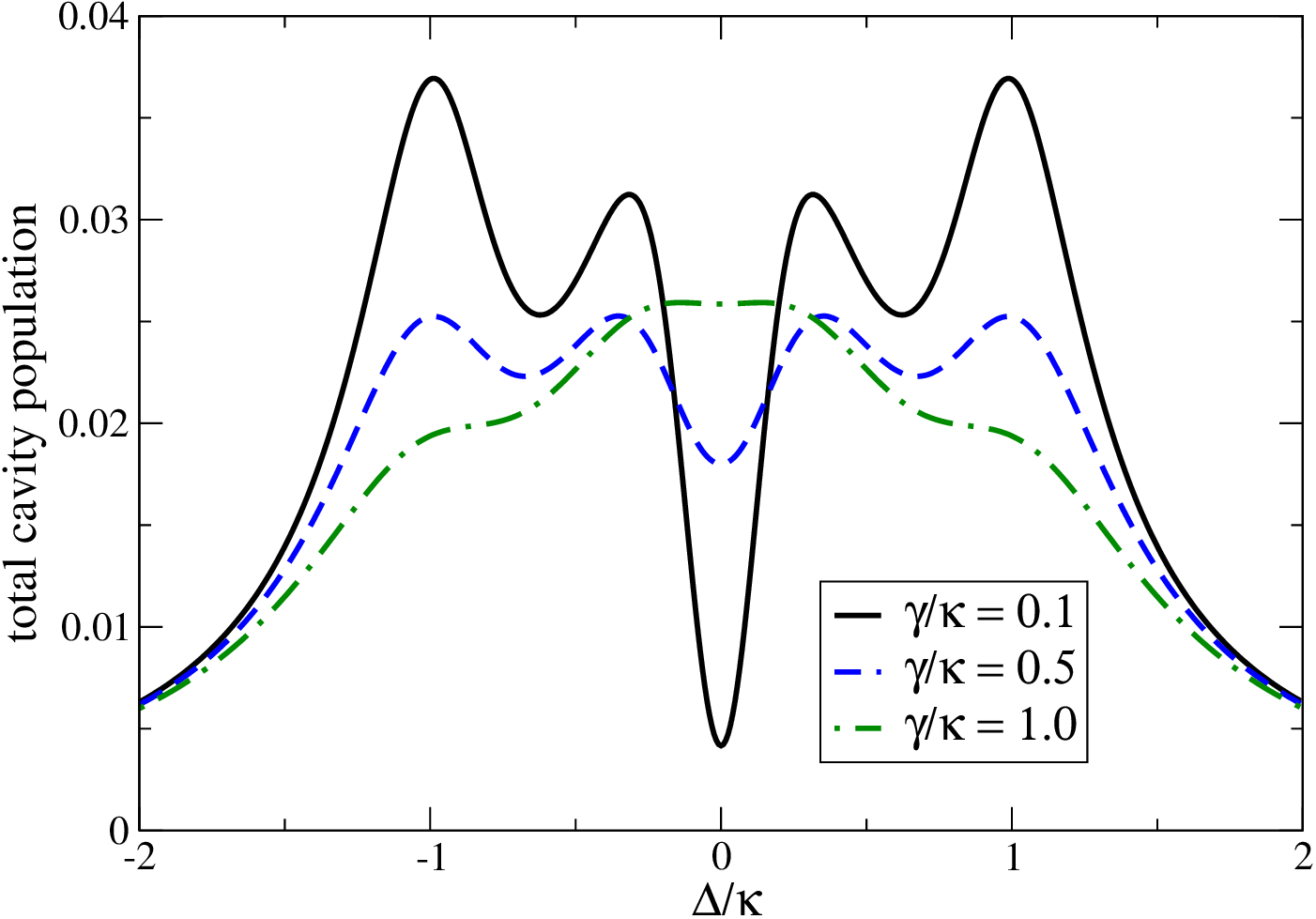}
    \caption{ 
    As \fig{FIG:DSgamfinite} but here with two cavity mode, $M=2$. Plotted is the total cavity population, i.e. the sum from both modes.
    \label{FIG:DS2mode}
    }
  \end{center}
\end{figure}
%%%%%%%%%%%%%%%%%%%%%%%%%%%%%%%%%%%%%%%%%%%%%%%%%%%%%%%%%%%%%%%%%

%%%%%%%%%%%%%%%%%%%%%%%%%%%%%%%%%%%%%%%%%%%%%%%%%%%%%%%%%
\subsection{Multi-mode case}
The existence of a dark state persists for arbitrary number of cavity modes $M$.
In the limit $\Delta_A \to 0$ (the value of $\Delta_C$ is unimportant), the state
\beq
  \ket{\Psi_D} =\mathcal{N}^{-1}
    \left\{
    \ket{\underline{0}}
    - \sum_k \frac{2\widetilde{\eta}_k}{\lambda_k}\ket{\widetilde{A}_k}
  \right\}
  ,
\eeq
with the norm $\mathcal{N}^2 = 1 +4\sum_k \frac{|\widetilde{\eta}_k|^2}{\lambda_k^2} $
is once again both the zero-eigenvalue eigenstate of Hamiltonian, \eq{Hsplit}, and the stationary state of the master equation.  \fig{FIG:DS2mode} shows the total cavity population
\beq
  \sum_{m=1}^M \ew{ a_m^\dag a_m} = \sum_{m=1}^M \ew{ \op{C_m}{C_m}}
  = \sum_{m=1}^M \ew{ \op{\widetilde{C}_m}{\widetilde{C}_m}}
  \nonumber
  .
\eeq
for a particular two-mode example.  For small $\gamma/\kappa$, the four peaks at $\Delta = \pm \lambda_k$ are clearly visible, as is the anti-resonance at the origin originating from the dark state.  As $\gamma/\kappa$ increases, both the individual peaks and the anti-resonance wash out.

It is useful to consider two special cases.
In the completely symmetric case where $\lambda_i = \lambda$, $\eta_i =\eta^*_i = \eta$, the $M$-mode total cavity population with $\gamma=0$ can be obtained from the single-mode result, \eq{cavpopgam0}, by simply scaling the driving strength $\eta^2 \to M \eta^2$.  Thus moving to $M$-degenerate modes in this case does not significantly alter the form of the dark resonance or the criteria for its observation.  In the experiments of \citer{Wickenbrock2013} the splittings $\lambda_i$ were pseudo-degenerate, and so this is of relevance there.

The other interesting case is when the coupling matrix $G$ is such that there are a number of large splittings $\lambda_k\gg\kappa$ and a number of small ones $\lambda_k\lesssim\kappa$. An extreme version of this situation is when the atoms were localised in the cavity over distance that is small relative to any variation in the cavity field.  In this case, just one of the modes has a significant splitting with $\lambda_1 \sim g \sqrt{MN}$, whereas the remaining splittings are small.  The well-split mode is unaffected by the dark-state.  In the ideal case, the small splittings should set $\lambda_k\sim \kappa$, such that they are all affected by the dark resonance. The profile seen in  \fig{FIG:DS2mode} is a two-mode example of this type of profile, where both enhanced splittings due to collective effects and dark-anti-resonances are visible.
%

%%%%%%%%%%%%%%%%%%%%%%%%%%%%%%%%%%%%%%%%%%%%%%%%%%%%%%%%%%%%%%%%%%%%%%%%%
%%%%%%%%%%%%%%%%%%%%%       beyond        %%%%%%%%%%%%%%%%%%%%%%%%%%%%%%%
%%%%%%%%%%%%%%%%%%%%%%%%%%%%%%%%%%%%%%%%%%%%%%%%%%%%%%%%%%%%%%%%%%%%%%%%%
\section{Beyond weak excitation}

Milburn and Alsing \cite{Milburn2000} have considered the isolated system with $M=1$ and arbitrary $N$ with a symmetric coupling $g$, not restricted to small values.  For arbitrary values of $g$ below threshold, $ 4\eta/N g<1$ (see \citer{Milburn2000}), they showed that the ground state could be written as a product of a squeezed vacuum for the field and squeezed spin state for the atomic ensemble. The mean population of the cavity mode in this state was $\ew{a^\dag a} = -\frac{1}{4} \log \left\{1-\rb{\frac{4\eta}{Ng}}^2\right\}$. 
Therefore, away from the weak-driving limit, the system is not completely dark. 
However, provided that the driving is weak enough that $ 4\eta/N g \ll 1$, the cavity population in the stationary state of the system will be
\beq
  \ew{a^\dag a} \approx \rb{\frac{2\eta}{Ng}}^2
  .
\eeq
This can be made as small as required, either by making the driving weaker or by increasing $N$.  As both these factors can be controlled, the finite emission coming from deviation from the weak-excitation limit can be made small with respect to the other source of imperfect darkness, spontaneous emission.

%%%%%%%%%%%%%%%%%%%%%%%%%%%%%%%%%%%%%%%%%%%%%%%%%%%%%%%%%%%%%%%%%%%%%%%%%
%%%%%%%%%%%%%%%%%%%%%      Conclusions    %%%%%%%%%%%%%%%%%%%%%%%%%%%%%%%
%%%%%%%%%%%%%%%%%%%%%%%%%%%%%%%%%%%%%%%%%%%%%%%%%%%%%%%%%%%%%%%%%%%%%%%%%
\section{Conclusions}

We have considered the occurrence of dark states in an $N$-atom $M$-mode Jaynes-Cummings system in the weak driving limit.  If each of the effective mode-splittings satisfy $\lambda_k \gg \sqrt{2\gamma \kappa}$ then emission at zero detuning, $\Delta_A=0$, will be suppressed. However, only if at least one mode-splitting is of the order of the cavity-broadening $\lambda_k\sim\kappa$ will a pronounced anti-resonance, characteristic of dark states be observable.

It is perhaps interesting to note that when, as here, the cavity field is driven, it is the emission from the cavity that is suppressed. However, if the atom(s) are driven instead, it is their fluorescence that is suppressed \cite{Cirac1993,Rice1996,Zippilli2004}.

%%%%%%%%%%%%%%%%%%%%%%%%%%%%%%%%%%%%%%%%%%%%%%%%%%%%%%%%%%%%%%%%%%%%%%%%%
\begin{acknowledgments}
I am grateful to F.~Renzoni, G.~R.~M.~Robb, and A.~Wickenbrock for useful discussions.

\end{acknowledgments}

%%%%%%%%%%%%%%%%%%%%%%%%%%%%%%%%%%%%%%%%%%%%%%%%%%%%%%%%%%%%%%%%%%%%%%%%%
%%%%%%%%%%%%%%%%%%%%%      REFERENCES     %%%%%%%%%%%%%%%%%%%%%%%%%%%%%%%
%%%%%%%%%%%%%%%%%%%%%%%%%%%%%%%%%%%%%%%%%%%%%%%%%%%%%%%%%%%%%%%%%%%%%%%%%
%Merlin.mbs v4.21 2009-07-09.
%

\end{document}